%% file: FWD-11-003_temp.tex
\pdfoutput=1

\documentclass[11pt,twoside,a4paper,cmspaper,final,collab]{cms-tdr}
\def\svnVersion{187809}\def\cmsCernNo{2013-012}\def\cmsCernDate{\today}\def\cmsMessage{Submitted to the Journal of High Energy Physics}

\begin{document}\cmsNoteHeader{FWD-11-003}

\hyphenation{had-ron-i-za-tion}
\hyphenation{cal-or-i-me-ter}
\hyphenation{de-vices}

\RCS$Revision: 177202 $
\RCS$HeadURL: svn+ssh://svn.cern.ch/reps/tdr2/papers/FWD-11-003/trunk/FWD-11-003.tex $
\RCS$Id: FWD-11-003.tex 177202 2013-03-19 12:06:01Z roland $
\cmsNoteHeader{FWD-11-003} 
\title{Study of the underlying event at forward rapidity  in pp collisions at \texorpdfstring{$\sqrt{s} = 0.9$, 2.76, and 7\TeV}{sqrt(s) = 0.9, 2.76, and 7 TeV}}

\date{\today}

\abstract{
The underlying event activity in proton-proton collisions at forward pseudorapidity $(-6.6 < \eta < -5.2)$ is studied with the CMS detector at the LHC, using a novel observable:  the ratio of the forward energy density, $\rd{}E/\rd\eta$, for events with a charged-particle jet produced at central pseudorapidity ($|\eta^\text{jet}| < 2$) to the forward energy density for inclusive events.  This forward energy density ratio is measured as a function of the central jet transverse momentum, $\pt$, at three different pp centre-of-mass energies ($\sqrt{s} = 0.9$, 2.76, and 7\TeV).  In addition, the $\sqrt{s}$ evolution of the forward energy density is studied in inclusive events and in events with a central jet. The results are compared to those of Monte Carlo event generators for pp collisions and are discussed in terms of the underlying event. Whereas the dependence of the forward energy density ratio on jet $\pt$ at each $\sqrt{s}$ separately can be well reproduced by some models, all models fail to simultaneously describe the increase of the forward energy density with $\sqrt{s}$ in both inclusive events and in events with a central jet.
}

\hypersetup{%
pdfauthor={CMS Collaboration},%
pdftitle={Study of the underlying event at forward rapidity in pp collisions at sqrt(s) = 0.9, 2.76, and 7 TeV},%
pdfsubject={underlying event},%
pdfkeywords={underlying event, multiple-parton interactions, forward energy flow}}

\maketitle 

\section{Introduction}
\label{sec:introduction}

Particle production in soft, nondiffractive inelastic collisions between hadrons is characterized by a particle density that is uniform in rapidity within a rapidity range proportional to $\ln s$, where $\sqrt{s}$ is the centre-of-mass energy of the collision (see, e.g.\@ \cite{Khachatryan:2010xs,Khachatryan:2010us}). The particle density and the average momentum per particle slowly increase with $s$, and, as a consequence, the energy density per unit of rapidity is expected to show an approximately~logarithmic~increase~with~$s$~\cite{kittel2005}.

This picture changes when a hard scattering occurs in the collision, resulting in two back-to-back, large transverse-momentum ($\pt$) jets. These are accompanied by hadronic activity due to initial- and final-state parton showers.  In the commonly used DGLAP approach \cite{gribov1972,lipatov1975,altarelli1977,dokshitzer1977}, the transverse momentum $\kt$ of these parton showers increases as their rapidity approaches the rapidity of the partons emerging from the hard interaction.  Alternative models for parton dynamics, such as BFKL \cite{kuraev1976,kuraev1977,balitsky1978} or CCFM \cite{ciafaloni1988,catani1990,catani1990-2,marchesini1995}, however, also allow large-$\kt$ parton emissions far away from the hard scatter, thus yielding a larger energy density at rapidities well separated from the high-$\pt$ jets.  Additionally, the incoming particle remnants may re-scatter and fragment, thereby producing a final state similar to that of soft collisions.  Parton showers and remnant interactions form the so-called underlying event.

Previous studies \cite{cms2010,atlas2011,atlas2011-2,cms2011,alice2011} typically separate hadronic activity due to the underlying event from activity resulting from the hard scattering by dividing the azimuthal plane into the so-called toward, transverse, and away regions with respect to the direction of the highest-$\pt$ jet.  The hadronic activity in the transverse region is then assumed to be dominated by the underlying event, while the toward and away regions are also populated by the jets. A complementary method, followed in this paper, consists of studying the hadronic activity in a region far away in rapidity from the hard-scattering products.  The toward, transverse, and away regions are then all dominated by the underlying event.

In the present paper, the underlying event activity is studied at forward pseudorapidity ($-6.6 < \eta < -5.2$) in a novel way by measuring the ratio of the forward energy density per unit of pseudorapidity for events with a charged-particle jet produced at central pseudorapidity ($|\eta^\text{jet}| < 2$) to the forward energy density for inclusive, dominantly nondiffractive, events.  This energy density ratio is measured as a function of the jet transverse momentum at three different proton-proton centre-of-mass energies ($\sqrt{s} = 0.9$, 2.76, and 7\TeV).  In addition, the relative increase of the forward energy density as a function of centre-of-mass energy is presented for inclusive events and for events with a central charged-particle jet.   This extends the study of the forward energy density in the pseudorapidity range $3 < |\eta| < 5$ published in \cite{Chatrchyan:2011wm} to a previously unexplored region.

The paper is structured as follows. A discussion of the phenomenology of the underlying event is given in Section~\ref{sec:phenomenology}. Monte Carlo (MC) simulation programs used to correct data for detector effects and to compare models to corrected data are discussed in Section~\ref{sec:mc}.  Section \ref{sec:detector} gives a short description of the CMS detector.
The analysis is discussed in Sections~\ref{sec:trigger} and~\ref{sec:datacorrection}.  Section~\ref{sec:systematics} describes the investigation of systematic uncertainties.  Results are discussed in Section~\ref{sec:results} and  a summary is given in Section~\ref{sec:conclusion}.

\section{Phenomenology of the underlying event}
\label{sec:phenomenology}

One theoretical framework used to describe the underlying event is the multiple-parton interaction (MPI) model, which assumes that parton interactions occur in addition to the primary hard scattering.  These additional interactions are softer than the primary one, but still perturbatively calculable.

The requirement of jets in the final state selects, on average, collisions with a smaller impact parameter \cite{sjostrand1987,frankfurt2011}.  In the MPI model as implemented in \PYTHIA6 \cite{sjostrand2006}, this correlation is realised by a suppression factor of low-$\pt$ parton interactions at small impact parameter. Such central collisions have a larger overlap of the matter distributions of the colliding hadrons and are therefore more likely to have many parton interactions.  The comparison of particle and energy densities between events with hard jets in the final state and inclusive events thus yields information on underlying events with many parton interactions relative to those with few of them.

\begin{figure}[t]
\begin{center}
\includegraphics[width=0.6\textwidth]{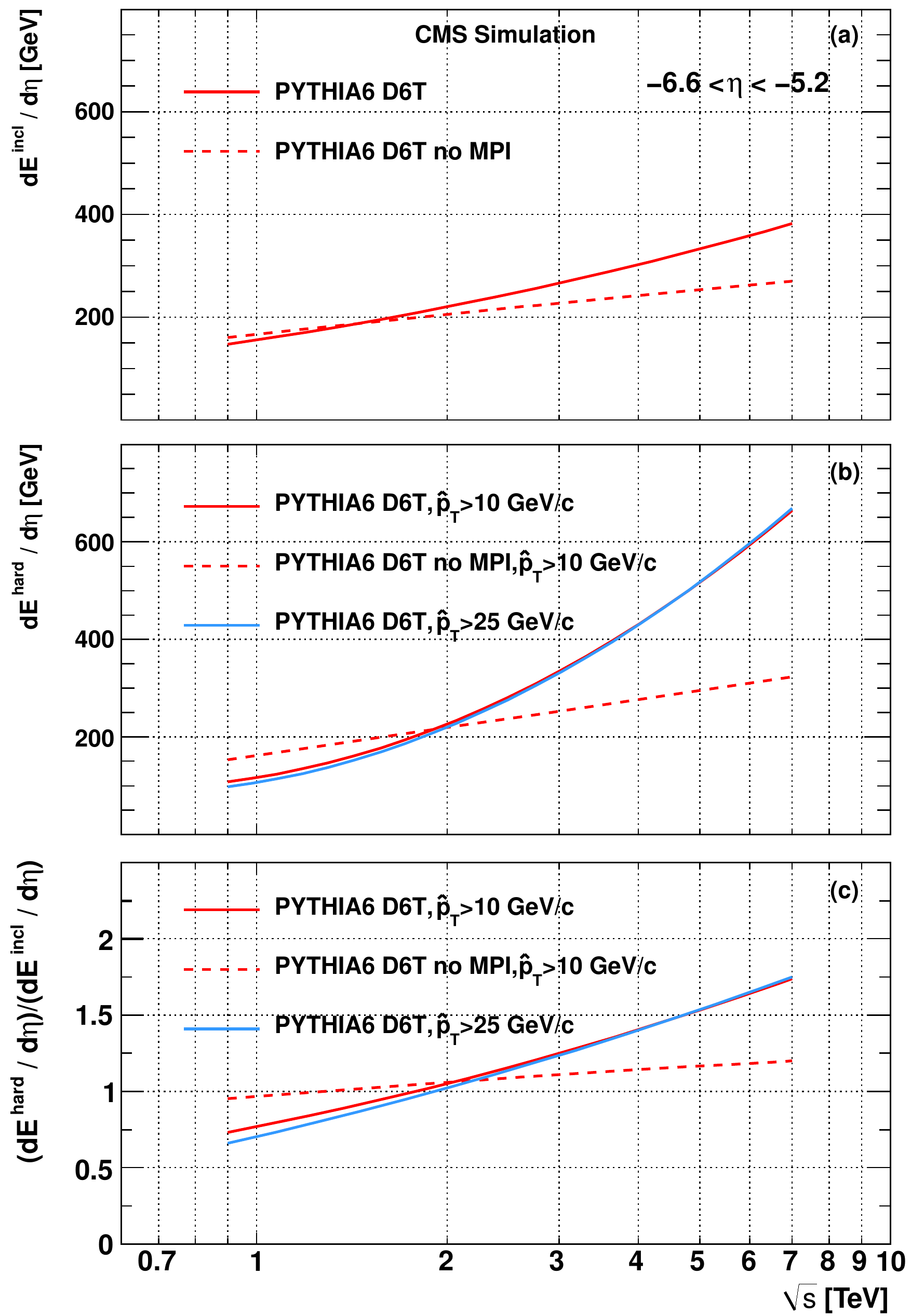}
\end{center}
\caption{The energy density $dE/d\eta$ in the pseudorapidity region $-6.6 < \eta < -5.2$, obtained with \PYTHIA6 D6T,  is plotted as a function of $\sqrt{s}$, for  inclusive, nondiffractive events (a) and for events with a central ($|\eta| < 2$) hard parton interaction with transverse momentum transfer, $\hat{p}_{\rm T}$, above a given threshold (b).  The ratios of the plots in (a) and (b) are shown in (c).}
\label{fig:genstudy}
\end{figure}

Figure \ref{fig:genstudy} shows the result of a simulation based on the D6T underlying event tune \cite{field2008,field2009} of the \PYTHIA6 generator. Although it is not the best tune to describe early measurements of the underlying event activity at the Large Hadron Collider (LHC), this tune is used here because it yields a large number of MPIs, which results in an enhanced effect on the forward energy density. Other tunes show a similar, albeit somewhat reduced, behaviour. Figure \ref{fig:genstudy}a shows the energy density, $\rd{}E/\rd\eta$, for $-6.6 < \eta < -5.2$,  as a function of $\sqrt{s}$ for inclusive events. Figure \ref{fig:genstudy}b shows the energy density for events with a central ($|\eta| < 2$) hard parton interaction with transverse momentum transfer, $\hat{p}_{\rm T}$, above 10 or 25\GeVc. Finally, Fig.\@~\ref{fig:genstudy}c shows the ratio of these two distributions, henceforward called the ``hard-to-inclusive forward energy ratio''.

It can be seen that the energy density in inclusive events is only slightly affected by the presence of MPIs.  This is not the case in events with a hard parton interaction at large $\sqrt{s}$, where a large increase of the energy density is predicted when including MPIs.  Moreover, this increase is roughly independent of $\hat{p}_{\rm T}$, indicating that collisions are already central for $\hat{p}_{\rm T} > 10$\GeVc.  Finally, the hard-to-inclusive forward energy ratio would be close to unity in the absence of MPIs.  With MPIs, however, the ratio is substantially higher than 1 at large $\sqrt{s}$, while it drops below 1 at small $\sqrt{s}$.  This last observation points to a depletion of the energy of the proton remnant in events with hard central jets.  Indeed, at $\sqrt{s} = 0.9 {\rm\\TeV}$, the proton remnant has a rapidity $y = \ln\left({\sqrt{s}/m_p}\right) \approx 7$ where $m_p$ is the proton rest mass. At this centre-of-mass energy, the energy density in the considered pseudorapidity range is thus sensitive to the details of remnant fragmentation.

\section{Monte Carlo models}
\label{sec:mc}

In this section, the main features of the Monte Carlo models used in the analysis are presented, with emphasis on the implementation and tuning of the underlying event.

Several tunes of the \PYTHIA6 (version 6.424) \cite{sjostrand2006} and \PYTHIA8 (version 8.145) \cite{sjostrand2008} event generators are used, each providing a different description of the underlying event in nondiffractive interactions: D6T \cite{field2008,field2009}, Z2 \cite{cms2011} and Z2* for \PYTHIA6 and 4C  \cite{corke2010} for \PYTHIA8.  The parameter settings in D6T were determined from the Tevatron data, while the other tunes were determined from the LHC data on inclusive and underlying event properties at central pseudorapidity.    The more recent \PYTHIA6 Z2 and Z2* tunes, as well as \PYTHIA8, use a new model \cite{skands2007} where multiple-parton interactions are interleaved with parton showering.  The Z2 and Z2* tunes are derived from the Z1 tune \cite{field2010}, which uses the CTEQ5L parton distribution set, whereas Z2 and Z2* adopt CTEQ6L.  The Z2* tune is the result of retuning the \PYTHIA6 parameters PARP(82) and PARP(90) by means of the automated \textsc{Professor} tool \cite{buckley2010}, yielding PARP(82)=1.921 and PARP(90)=0.227.  The results of this study are also compared to predictions obtained with \PYTHIA6, tune Z2*, with multiple-parton interactions switched off.  \PYTHIA8 is used with tune 4C, based on the early LHC data.  Parton showers in \PYTHIA are modelled according to the DGLAP prescription.

The \HERWIG{}++  (version 2.5) \cite{bahr2008} MC event generator, with a recent tune to LHC data (UE-EE-3C \cite{Gieseke:2011na}),   is used for comparison to data.  The evolution of the parton distribution functions with momentum scale in \HERWIG{}++ is also driven by the DGLAP equations.

In contrast to \PYTHIA and \HERWIG{}++, \textsc{cascade} \cite{Jung:2000hk,Jung:2010si} is based on the CCFM evolution equation for the initial-state cascade, supplemented with off-shell matrix elements for the hard scattering. Multiple-parton interactions are not implemented in \textsc{cascade}.

The \textsc{dipsy}  generator \cite{Flensburg:2011kk} is based on a dipole picture of BFKL evolution.  It includes multiple dipole interactions, with parameters tuned as described in \cite{Flensburg:2011kk}, and can be used to predict nondiffractive final states. In the present implementation, however, quarks are not included in the evolution. The treatment of the proton remnant and valence quark structure is therefore simplistic, and predictions for the structure of the final state in the very forward region are somewhat uncertain.

Finally, data are also compared to the predictions of  Monte Carlo pp event generators used in cosmic-ray physics \cite{d'enterria2011}. The generators \textsc{epos}1.99 \cite{werner2006} , \textsc{QGSJet}II \cite{ostapchenko2011},
and \textsc{sybill} 2.1 \cite{ahn2009} are considered (for an overview, see \cite{Ostapchenko:2006aa}).  In general, these models describe the soft component in terms of the exchange of virtual quasi-particle states, as in Gribov's reggeon field theory \cite{gribov1968}, with multi-pomeron exchanges accounting for MPI effects. At higher energies and scales, the interaction is described by perturbative Quantum Chromodynamics (QCD) (with DGLAP evolution).  These models also include non-linear parton effects, either by including pomeron-pomeron interactions, as in \textsc{QGSJet} and \textsc{EPOS},  or by means of a parton saturation approach, as in \textsc{sybill}. These cosmic ray models were not tuned to LHC data.

\section{The CMS detector}
\label{sec:detector}

The central feature of the Compact Muon Solenoid (CMS) apparatus is a superconducting solenoid of 6\unit{m} internal diameter. Within the field volume are a silicon pixel and strip tracker, a crystal electromagnetic calorimeter  and a brass/scintillator hadron calorimeter.  Muons are measured in gas-ionization detectors embedded in the steel flux return yoke. In addition to the barrel and endcap detectors, CMS has extensive forward calorimetry.

The CMS experiment uses a right-handed coordinate system, with the origin at the nominal interaction point, the $x$-axis pointing to the center of the LHC ring, the $y$-axis pointing up (perpendicular to the plane of the LHC ring), and the $z$-axis along the anticlockwise-beam direction. The polar angle $\theta$ is measured from the positive $z$-axis and the azimuthal angle $\phi$ is measured in the $x$-$y$ plane.  Pseudorapidity is defined as $\eta = -\ln\tan(\theta/2)$ and approximates true rapidity $y = \ln\frac{E+p_z}{E-p_z}$. Pseudorapidity equals rapidity for massless particles.

The tracker measures charged particles within the pseudorapidity range $|\eta| < 2.5$. It consists of 1\,440 silicon pixel and 15\,148 silicon strip detector modules and is located in the 3.8~T field of the superconducting solenoid. It provides an impact parameter resolution of ${\sim}15\mum$ and a transverse momentum resolution of about 1.5\% for 100\GeVc particles.

The hadronic forward (HF) calorimeters cover the region $2.9< |\eta| <5.2$.  They consist of iron absorbers and embedded radiation-hard quartz fibres read out by radiation-hard photomultiplier tubes (PMTs).  Calorimeter cells are formed by grouping bundles of fibres.  Clusters of these cells form a calorimeter tower.  There are 13 towers in $\eta$, each with a size $\Delta\eta \approx  0.175$, except for the lowest- and highest-$|\eta|$ towers with $\Delta\eta\approx 0.1$ and $\Delta\eta \approx 0.3$, respectively. The azimuthal segmentation $\Delta\phi$ of all towers is $10^\circ$, except for the ones at highest-$|\eta|$, which have $\Delta\phi = 20^\circ$.

More forward angles, $-6.6 < \eta < -5.2$, are covered by the Centauro And Strange Object Research (CASTOR) calorimeter, which is located only on the negative-$z$ side of CMS, at 14.37\unit{m} from the interaction point. The calorimeter is segmented in 16 $\phi$-sectors and 14 $z$-modules, corresponding to a total of 224 cells.  Each cell consists of 5 quartz plates of 4\unit{mm} thickness (2\unit{mm} for the electromagnetic modules) embedded in  5 tungsten absorber plates of 10\unit{mm} thickness (5\unit{mm} for the electromagnetic modules), with $45^\circ$  inclination with respect to the beam axis.  Air core light guides provide a fast collection of the \v{C}erenkov light to fine-mesh PMTs \cite{h11998}, which can operate in magnetic fields up to 0.5\unit{T} if the field direction is within ${\pm}45^\circ$ with respect to the PMT axis \cite{goettlicher2010}.   The first two modules, which have an absorber thickness half of that of the other modules, are used to detect electromagnetic showers.  The full calorimeter has a depth of 10.5 interaction lengths. However, the responses of PMTs reading out modules 6 to 8 are affected by the fringe field of the CMS solenoid. Therefore, only the 5 front modules in a $\phi$ sector are used, with the signals from the cells in a $\phi$ sector grouped into a so-called tower.  These 5 modules correspond to 3.23 interaction lengths and detect 80\% of the hadronic showers in inclusive events, on average.   Test beam measurements with a full-length CASTOR prototype \cite{goettlicher2010} were used to validate the simulation of the detector response.

Including the HF and CASTOR forward calorimeters, the CMS detector covers the range $-6.6 < \eta < +5.2$.

For the online selection of events, the CMS trigger system is used, together with two elements of the CMS detector monitoring system: the beam scintillation counters (BSC) and the Beam Pick-up Timing for the eXperiments (BPTX). The BSCs cover the region $3.23 < |\eta| < 4.65$. The BPTX devices are located around the beampipe at a distance of ${\pm}175\unit{m}$ on both side of the IP and are designed to provide precise information on the bunch structure and timing of the incoming beam.

A more detailed description of the CMS detector can be found in \cite{cms2008}.

\section{Event selection and reconstruction}
\label{sec:trigger}

This analysis is based on data collected in 2010 and 2011 at $\sqrt{s} = 0.9$, 2.76, and $7\TeV$, corresponding to integrated luminosities of $0.19\nbinv$, $0.30\nbinv$, and $0.12\nbinv$, respectively.  Runs are selected by requiring that the relevant components of the CMS detector were fully functional.  The average number of collisions per bunch crossing, inferred from the instantaneous luminosity and the total inelastic cross section, in each of the runs considered for this analysis is 0.017, 0.22, and 0.12 at $\sqrt{s} = 0.9$, 2.76, and 7\TeV, respectively.

The CMS data acquisition was triggered by the presence of hits in both BSC detectors, for the 0.9 and 7\TeV data sample, or hits in either of the BSC detectors, for the 2.76\TeV data sample. Standard CMS algorithms to remove beam halo events are applied.

A sample of inclusive nondiffractive events is selected offline, with minimal bias, by requiring exactly one primary vertex, at least one HF tower with energy larger than 4\GeV in the pseudorapidity range of each BSC detector, and at least one CASTOR tower with energy above 1.5\GeV.   The numbers of selected minimum-bias events are 4.7, 9.8, and 4.6 million at $\sqrt{s} = 0.9$, 2.76, and $7\TeV$, respectively.

Track jets are reconstructed with the anti-$\kt$ algorithm \cite{cacciari2008}  with a size parameter of 0.5, applied to tracks fitted to a primary vertex and with transverse momentum of at least 0.3\GeVc.  The leading track jet with $\pt > 1\GeVc$ and $|\eta^\text{jet}| < 2$ defines the hard scale in the event. An advantage of using  track jets is that they are  experimentally well-defined objects. No attempt is made to correct to the corresponding parton-level objects, as this would result in additional model uncertainties. Moreover, track jets are much better correlated in energy and direction to partons than the highest-$\pt$ track. Finally, in the few $\GeVc$ region, the $\pt$ of a track jet is better determined than the $\pt$ of calorimeter-based jets, which suffer from poor energy resolution at low $\pt$.

The total charge collected by the PMTs of the 5 front $z$-modules of the CASTOR calorimeter is used to measure the energy deposited in the CASTOR $\eta$ range.  The response of individual CASTOR cells is equalized by using a sample of beam halo muon events.  An absolute calibration factor of $0.015$\GeV/fC, with an uncertainty of ${\pm}20\%$, is obtained from an extrapolation of the $\eta$ dependence of the energy density in HF to the CASTOR acceptance region and is found to be consistent with the results of test beam measurements.  The energy ratios presented in this analysis, however, do not depend on the absolute calibration and are only marginally affected by the relative inter-calibration of channels.

\section{Data correction}
\label{sec:datacorrection}

In order to be able to compare to theoretical predictions, the data are corrected for various detector effects, including trigger efficiency, event selection efficiency, energy reconstruction efficiency in CASTOR and smearing effects in track jet $\pt$.  Except for the trigger efficiency correction, which is extracted directly from data, corrected results are obtained by means of a simulation of the CMS detector based on \GEANTfour \cite{agostinelli2003,allison2006}.

The trigger conditions and event selection criteria outlined in Section~\ref{sec:trigger} are chosen to select a sample of dominantly nondiffractive events. However,  high-mass diffractive dissociation events, covering the full detector and having a large rapidity gap outside the acceptance, remain in the data sample.  A precise definition of the phase space, at the level of stable particles, for which corrected results are presented is obtained as follows.

The collection of stable (lifetime $\tau > 10^{-12}\unit{s}$) final-state particles is divided into two systems, $X$ and $Y$, based on the mean rapidity of the two particles separated by the largest rapidity gap in the event.  All particles on the negative side of the largest gap are assigned to the system $X$, while the particles on the positive side are assigned to the system $Y$ \cite{h11997}. The invariant masses, $M_\text{X}$ and $M_\text{Y}$, of each system are calculated by using the four-momenta of the individual particles; their ratios to the centre-of-mass energy, $\xi_\text{X}$, $\xi_\text{Y}$, and $\xi_\text{DD}$, are defined as follows:
\begin{equation}
\xi_\text{X} = \frac{M_\text{X}^2}{s}, \qquad \xi_\text{Y} = \frac{M_\text{Y}^2}{s}, \qquad \xi_\text{DD} = \frac{M_\text{X}^2 M_\text{Y}^2}{m_p^2\, s},
\end{equation}
where $m_p$ is the proton rest mass and the subscript DD refers to double diffractive dissociation.  These Lorentz-invariant variables are well-defined for any type of events.  In the case of large rapidity gap events, they are related to the size of the rapidity gap via $\Delta y \simeq \ln 1/\xi$.

The phase space remaining for events with a large rapidity gap, after applying detector-level selection criteria, can then be quantified at the stable-particle level by appropriate limits on $\xi_\text{X}$, $\xi_\text{Y}$, and $\xi_\text{DD}$.  These acceptance limits are obtained from a dedicated study based on \PYTHIA6 (tune Z2*) using fully simulated events and are tabulated in Table\@~\ref{tab:xilimits}.  An event is selected at the stable-particle level if {\em any} of $\xi_\text{X}$, $\xi_\text{Y}$, or $\xi_\text{DD}$ is larger than the respective limit.  Because the detector acceptance changes with centre-of-mass energy, different thresholds are used at $\sqrt{s} =$ 0.9, 2.76, and 7\TeV.  In all cases, however, the selection applied ensures that there are no large gaps inside the detector acceptance. Adapting the selected phase space dynamically to the detector acceptance results in a smaller correction of the data, and thus also in a smaller model dependence of the correction factors.

\begin{table}[t]
\caption{Acceptance limits on $\xi_\text{X}$, $\xi_\text{Y}$, and $\xi_\text{DD}$ used to define the phase space domain for which corrected results are presented.  These limits at the stable-particle level correspond to the phase space selected by detector level criteria.}\begin{center}
\begin{tabular}{|c|ccc|ccc|}
\hline
$\sqrt{s}$ (TeV) & $\xi_\text{X}^\text{min}$ & $\xi_\text{Y}^\text{min}$ & $\xi_\text{DD}^\text{min}$ \\ \hline
0.9 & 0.1 & 0.4 & 0.5  \\
2.76 & 0.07 & 0.2 & 0.5 \\
7  & 0.04 & 0.1 & 0.5 \\
\hline
\end{tabular}
\end{center}
\label{tab:xilimits}
\end{table}

Similarly to reconstructed track jets, jets at the stable-particle level are obtained by running an anti-$\kt$ algorithm, with a size parameter of 0.5, on stable charged particles with $\pt > 0.3\GeVc$ and $|\eta| < 2.5$.  Particle level jets are selected by requiring $\pt^\text{jet} > 1$\GeVc and $|\eta^\text{jet}| < 2$.

The trigger efficiency is determined from a sample of {\it zero bias\/} events.  Zero bias events are triggered by the BPTX devices, which require to have filled bunches crossing each other in the CMS interaction point. The efficiency of the trigger used for the collection of minimum-bias events is determined as the fraction of events that have been triggered in a sample of offline
selected zero bias events.  The overall efficiency for triggering on the coincidence of a hit in both BSCs is 96.5\% (98.4\%) at $\sqrt{s} =$ 0.9 (7)\TeV.   For $\sqrt{s} =$ 2.76\TeV, where a trigger based on a hit in either BSC is used, the overall trigger efficiency is 99.9\% and no further correction is applied.
The efficiency at $\sqrt{s} =$ 0.9 and 7\TeV is parameterized as a function of the energy measured by the HF calorimeters in the BSC pseudorapidity range.  To correct for the trigger inefficiency, a weight equal to the inverse of this  parameterized efficiency is applied to each observed event.

The results presented in Section~\ref{sec:results} are all based on ratios of energies reconstructed in CASTOR.  By measuring energy ratios, many systematic uncertainties, and, in particular, the absolute calibration uncertainty, cancel.  However, because of the noncompensating nature of the CASTOR calorimeter, the response may still vary with changing particle composition and energy spectrum.  The measured energy ratio is therefore corrected by a factor that depends on the measured central track jet $\pt$.  This correction, of at most 5\%, is obtained from the  \PYTHIA6 Z2 MC, reweighted as a function of the particle jet $\pt$ and of the total energy in CASTOR in order to maximize the agreement between data and simulation.

A further bin-by-bin correction is applied to account for migrations in track jet $\pt$.  The final correction factor applied to the data is the product of the two above-mentioned factors:
\begin{equation}
\frac{\rd{}E^\text{true}/\rd\eta (\pt{}_\text{jet}^\text{true})}{\rd{}E^\text{det}/\rd\eta(\pt{}_\text{jet}^\text{det})} =
\frac{\rd{}E^\text{true}/\rd\eta (\pt{}_\text{jet}^\text{true})}{\rd{}E^\text{true}/\rd\eta(\pt{}_\text{jet}^\text{det})} \times
\frac{\rd{}E^\text{true}/\rd\eta(\pt{}_\text{jet}^\text{det})}{\rd{}E^\text{det}/\rd\eta(\pt{}_\text{jet}^\text{det})},
\label{eq:cf}
\end{equation}
with the superscripts ``true'' and ``det'' referring to variables estimated at stable-particle level and detector level, respectively.  The first ratio on the right-hand-side of Eq.~(\ref{eq:cf}) corrects for migration in track jet $\pt$, while the second ratio is the correction factor applied to the energy measured in CASTOR. The overall correction factor varies between 0.96 and 1.06.

\section{Systematic uncertainties}
\label{sec:systematics}

Several sources of systematic uncertainties are investigated.  For each systematic effect, the full analysis is repeated and the deviations from the nominal result are added in quadrature to obtain the total systematic uncertainty. The following sources are considered and the corresponding uncertainties are summarized in Tables\@~\ref{tab:systematics} and \ref{tab:systematics2}:

\begin{table}[tp]
\caption{Systematic uncertainties on the hard-to-inclusive forward energy ratio for track jet $\pt> 10$\GeVc at different centre-of-mass energies.}
\begin{center}
\begin{tabular}{|l|ccc|}
\hline
Source of uncertainty & $\sqrt{s} =$ 0.9\TeV & $\sqrt{s} =$ 2.76\TeV & $\sqrt{s} =$ 7\TeV \\ \hline
CASTOR alignment            & 1.5\% & 2.9\% & 3.1\% \\
Noncompensation   & 1.1\% & 0.4\% & 0.6\% \\
Model dependence   & 3.0\% & 2.3\% & 1.3\% \\
Shower containment & 1.2\% & 1.4\% & 1.0\% \\
Noise suppression & 0.3\% & 0.2\% & 0.2\% \\ \hline
Total uncertainty      & 3.7\% & 4.0\% &  3.6\% \\
\hline
\end{tabular}
\end{center}
\label{tab:systematics}
\end{table}

\begin{table}[tp]
\caption{Systematic uncertainties on the relative energy density  vs.\@ $\sqrt{s}$ in inclusive events (incl.) and in events with a central charged-particle jet with $\pt > 10\GeVc$ (hard).}
\begin{center}
\begin{tabular}{|l|cccc|}
\hline
 Source of uncertainty & 0.9\TeV (incl.) & 0.9\TeV (hard) &  7\TeV (incl.) & 7\TeV (hard) \\ \hline
CASTOR alignment              & 8.0\% & 7.0\% & 2.5\% & 2.7\% \\
Non-compensation     & 0.5\% & 1.1\% & 0.1\% & 1.0\% \\
Model dependence      & 2.4\% & 3.6\% & 2.0\% & 2.2\% \\
Shower containment   & 1.2\% & 1.0\% & 1.3\% & 0.9\% \\
Noise suppression       & 0.6\% & 0.8\% & 0.6\% & 1.0\% \\ \hline
Total uncertainty        & 8.5\% & 8.0\% & 3.5\% & 3.9\% \\
\hline
\end{tabular}
\end{center}
\label{tab:systematics2}
\end{table}

\begin{itemize}

\item CASTOR alignment.\\ Sensors monitoring the position of CASTOR indicate that the detector moves by ${\sim}1\unit{cm}$ in the transverse plane when the CMS solenoid is switched on or off. The CASTOR alignment is therefore run period dependent. Some $\phi$ sectors move towards more central pseudorapidity and the range they cover changes to approximately $-6.3 < \eta < -5.13$, while corrected results are presented for the range $-6.6 < \eta < -5.2$.  A new correction factor is obtained by assuming a shift between the pseudorapidity range at the detector and the stable-particle level in the MC simulation equal to the displacement of the most affected sectors in data.  Corrected results are obtained as the average between the correction factors based on the nominal and the shifted position of CASTOR, with half the difference taken as systematic uncertainty.  In addition, for the study of the centre-of-mass energy dependence, a second systematic uncertainty is included in order to account for possible changes in the CASTOR position in runs at different $\sqrt{s}$. This is obtained from dedicated MC samples with CASTOR appropriately shifted.

\item Non-compensation. \\The CASTOR detector is a noncompensating calorimeter.  Measurements with a test beam setup \cite{goettlicher2010} have shown that the response to pions relative to that to electrons is ${\approx}50\%$. This ratio slowly increases with incoming particle energy, a behaviour described by the simulation.  The systematic uncertainty is obtained by scaling the response to hadronic showers in the simulation by the uncertainty (${\pm}5\%$) on the pion-to-electron response ratio obtained from the test beam measurement.

\item Model dependence. \\ Correction factors are obtained from MC simulation and may be model dependent.  The correction of the CASTOR energy ratio in particular is sensitive to the charged- to neutral-pion production ratio.  Therefore, different response factors are obtained from a generator level study based on the models used in the comparison with corrected results.  The response factors are defined as the sum of the electromagnetic energy and 50\% of the hadonic energy divided by the total energy deposited in CASTOR.  The largest relative variation in the response factors is taken as a systematic uncertainty on the correction factor.  In addition, fully simulated samples of \PYTHIA6 tune Z2 and \PYTHIA8  tune 4C are used to directly compute the model uncertainty on the correction factor.

\item Shower containment. \\ In this analysis only the 5 front modules of the CASTOR calorimeter are used.   In order to assess the systematic uncertainty due to the partial containment of the hadronic shower, the difference in the observed energy ratios obtained from simulations based on all 14 modules and those based on only the front 5 modules is taken as a contribution to the systematic uncertainty.

\item Noise suppression. \\ The noise threshold applied to CASTOR towers is varied by ${\pm}20\%$, reflecting the uncertainty in the absolute calibration factor.

\end{itemize}

The systematic uncertainty due to the effect of event overlays in one bunch crossing was found to be negligible.  Similarly, the systematic uncertainty resulting from the description of dead material in the detector model used in the simulation was found to be negligible for the relative measurements presented in this paper.

\section{Results}
\label{sec:results}

\begin{figure}[tp]
\begin{center}
\includegraphics[width=\textwidth,trim=0 70 50 0,clip]{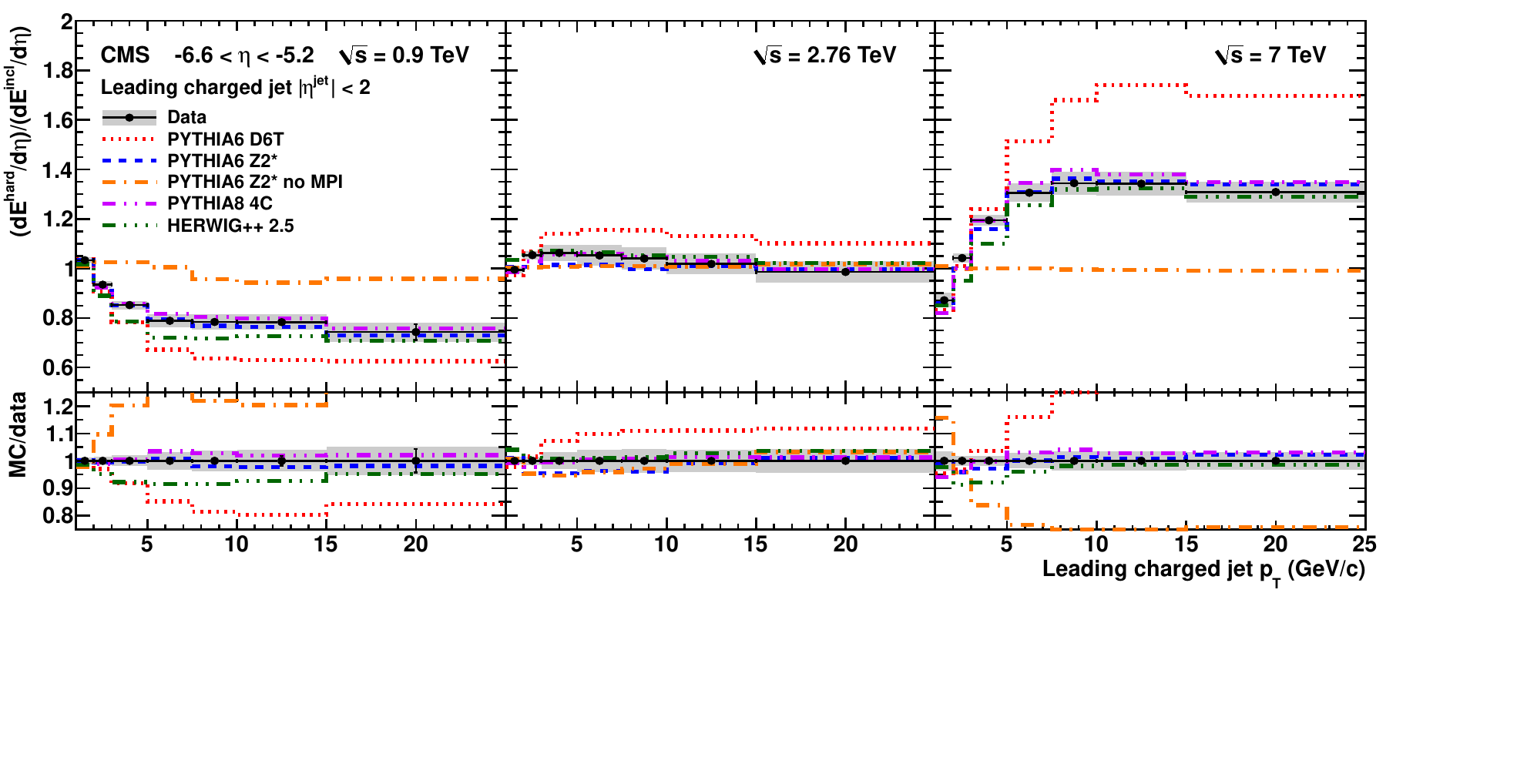}
\end{center}
\caption{Ratio of the energy deposited in the pseudorapidity range $-6.6 < \eta < -5.2$ for events with a charged-particle jet with $|\eta^\text{jet}| < 2$ with respect to the energy in inclusive events, as a function of the jet transverse momentum $\pt$ for $\sqrt{s} =$ 0.9 (left), 2.76 (middle), and 7\TeV (right).  Corrected results are compared to the \PYTHIA and \HERWIG{}++ MC models. Error bars indicate the statistical uncertainty on the data points, while the grey band represents the statistical and systematic uncertainties added in quadrature.}
\label{fig:ratiovspt}
\end{figure}

\begin{figure}[tp]
\begin{center}
\includegraphics[width=\textwidth,trim=0 70 50 0,clip]{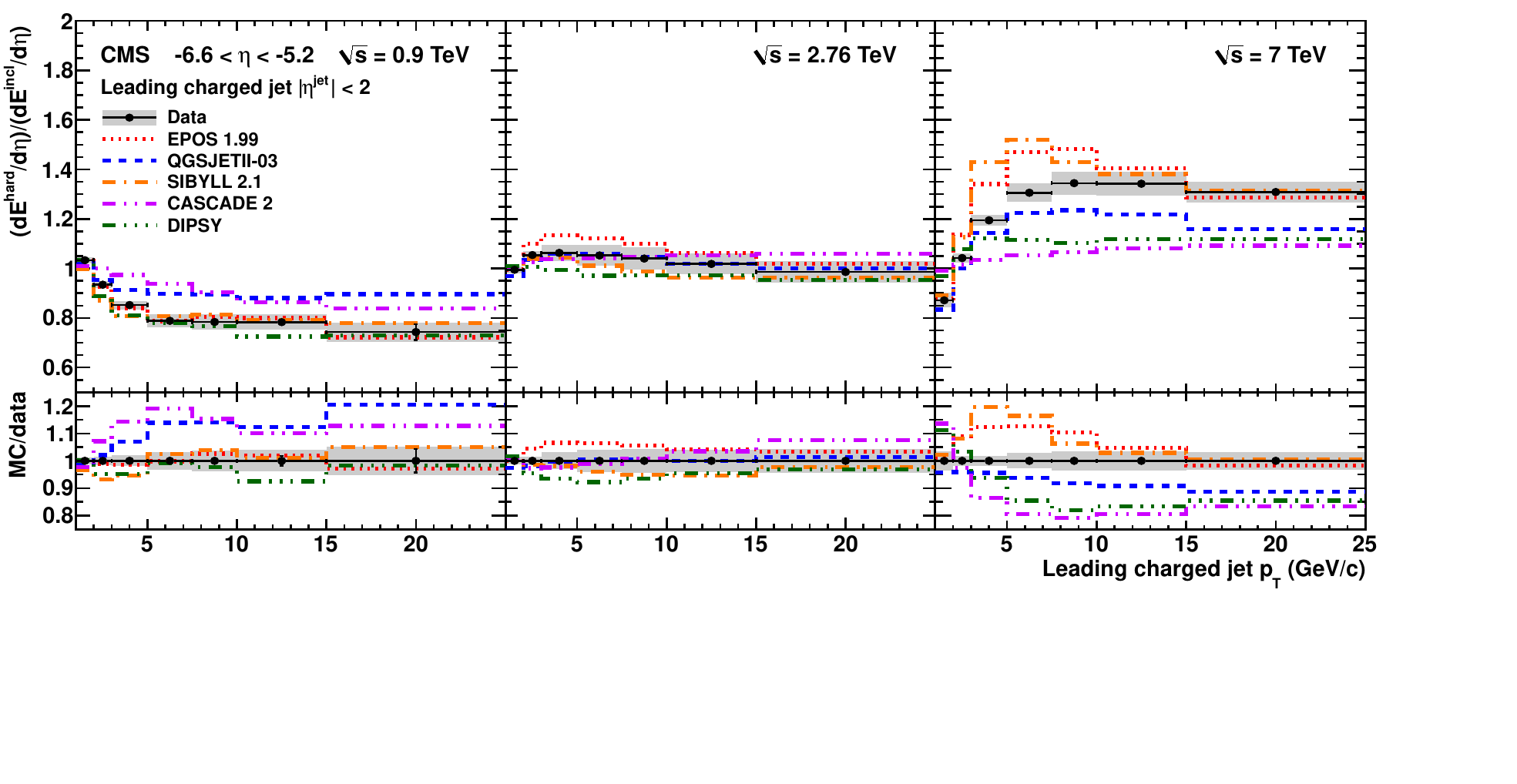}
\end{center}
\caption{Ratio of the energy deposited in the pseudorapidity range $-6.6 < \eta < -5.2$ for events with a charged-particle jet with $|\eta^\text{jet}| < 2$ with respect to the energy in inclusive events, as a function of the jet transverse momentum $\pt$ for $\sqrt{s} =$ 0.9 (left), 2.76 (middle), and 7\TeV (right).  Corrected results are compared to MC models used in cosmic ray physics and to \textsc{cascade} and \textsc{dipsy}. Error bars indicate the statistical uncertainty on the data points, while the grey band represents the statistical and systematic uncertainties added in quadrature.}
\label{fig:ratiovspt_cr}
\end{figure}

All the data are fully corrected for detector effects as described in Section~\ref{sec:datacorrection}.  In particular, results are obtained for a sample of events dominated by nondiffractive collisions so that the energy density ratios are not biased by rapidity gaps in the CASTOR pseudorapidity range.

Figures~\ref{fig:ratiovspt} and \ref{fig:ratiovspt_cr} show the hard-to-inclusive forward energy ratios, defined as the energy deposited in the pseudorapidity range $-6.6 < \eta < -5.2$ in events with a charged-particle jet with $|\eta^\text{jet}| < 2$ divided by the energy deposited in inclusive, dominantly nondiffractive events, as a function of the jet transverse momentum $\pt$.  Both figures show the same data points, but compared to different models.

At $\sqrt{s} =7\TeV$, a fast increase is seen at low $\pt$ followed by a plateau above $\pt = 8$\GeVc.  In the framework of the MPI model for the underlying event, this can be understood from the relation between the impact parameter and the scale of the event, quantified by $\pt$.  As $\pt$ increases, the collisions become more central and the number of parton interactions increases. Above $\pt = 10\GeVc$, the collision is central and the underlying event activity saturates. The pre-LHC \PYTHIA6 tune D6T fails to describe the data, while \PYTHIA6 and \PYTHIA8 tunes fitted to LHC data on the underlying event at central rapidity agree with the data at forward rapidity within ${\pm}5\%$.  As expected, when MPIs are switched off, \PYTHIA predicts a forward energy density that is independent of the central jet $\pt$.  The \HERWIG{}++ 2.5 simulation with tune UE-EE-3C gives a slightly worse description of the data in the turn-on region, but is still within ${\pm}10\%$ of the measured points.  The \textsc{cascade} model, which does not simulate multiple-parton interactions, does not describe the data.  The discrepancy shows that the features observed in the data cannot be explained by the CCFM parton dynamics as implemented in this model.   The \textsc{dipsy} model, based on the BFKL dipole picture, and supplemented with multiple interactions between dipoles, however, also fails to describe the data.  Models used in cosmic ray physics, on the other hand, do describe the increase of the energy ratio as a function of $\pt$ reasonably well.  The \textsc{QGSJet}II-03 generator yields a ratio that is too low in the plateau region, while \textsc{Sibyll} 2.1, and \textsc{Epos} 1.99 overestimate the turn-on but converge on a very good description at large $\pt$.

At $\sqrt{s} =2.76\TeV$, the increase of the energy ratio with $\pt$ is much reduced.  This tendency is consistent with the result at $\sqrt{s} = 0.9\TeV$, where the ratio becomes less than unity. Here, the energy density in events with a central jet is thus lower than the energy density in inclusive events.  As discussed in Section \ref{sec:phenomenology}, this can be understood as a kinematic effect: the production of central hard jets, accompanied by higher underlying event activity (as seen in studies at central rapidity \cite{cms2011}), depletes the energy of the proton remnant, which at $\sqrt{s} =0.9\TeV$ fragments within the pseudorapidity region covered by CASTOR. This feature is roughly described by the models.  Again, the \PYTHIA6 D6T tune exhibits too strong an underlying event activity, even at $\sqrt{s} = 0.9\TeV$.  Other \PYTHIA tunes describe the data at $\sqrt{s} = 2.76\TeV$ and 0.9\TeV rather well. The \HERWIG{}++ 2.5 predictions lie slightly below the data at $\sqrt{s} = 0.9$\TeV, which indicates too strong an underlying event activity.  The \textsc{cascade} generator does not reproduce the data, while \textsc{dipsy} yields a reasonable description at these lower centre-of-mass energies.  Most of the  cosmic ray models describe the data well, with \textsc{QGSJet}II-03 again yielding slightly too low underlying event activity.

Overall, in this study, both the \PYTHIA6 Z2* and \PYTHIA8 4C tunes give a good  description of all data. This is in contrast with studies of the underlying event in the central region \cite{cms2011}, where  \PYTHIA6 Z2* gives an excellent description of the underlying event activity in the region transverse to the jet in azimuth (to which it was tuned), while \PYTHIA8 4C is too low.

\begin{figure}[tp]
\begin{center}
\includegraphics[width=\textwidth,trim=0 80 40 0,clip]{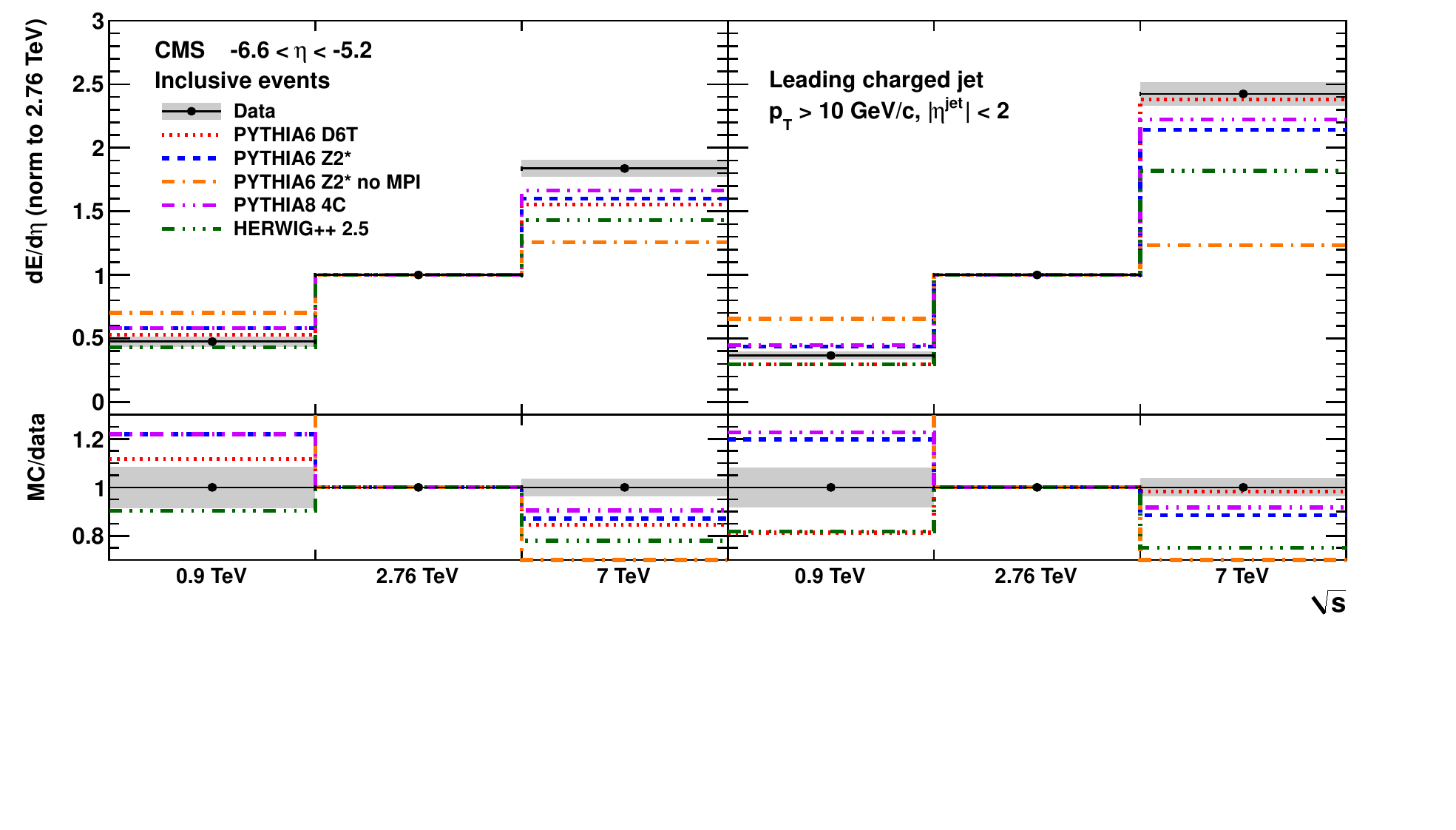}
\end{center}
\caption{Energy density in the pseudorapidity range $-6.6 < \eta < -5.2$ in inclusive events (left) and in events with a charged-particle jet in the range $|\eta^\text{jet}| < 2$ (right) as a function of $\sqrt{s}$, normalized to the energy density at $\sqrt{s} =2.76\TeV$.  The $\pt$ threshold used for jets is 10\GeVc at all centre-of-mass energies. Corrected results are compared to the \PYTHIA and \HERWIG{}++ MC models.  Statistical uncertainties are smaller than the marker size, while the grey band represents the statistical and systematic uncertainties added in quadrature.}
\label{fig:ratiovss}
\end{figure}

\begin{figure}[tp]
\begin{center}
\includegraphics[width=\textwidth,trim=0 80 40 0,clip]{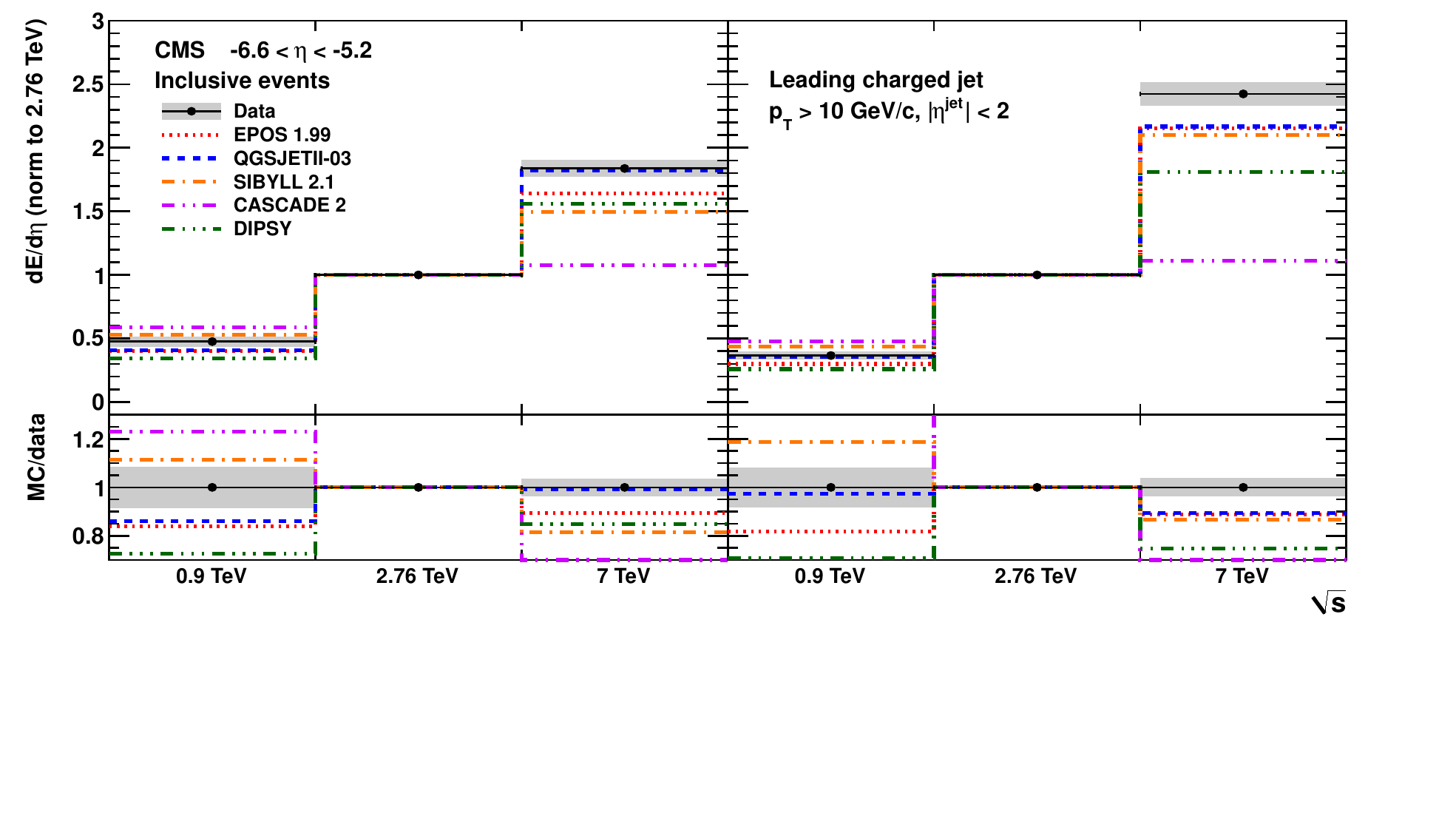}
\end{center}
\caption{Energy density in the pseudorapidity range $-6.6 < \eta < -5.2$ in inclusive events (left) and in events with a charged-particle jet in the range $|\eta^\text{jet}| < 2$ (right) as a function of $\sqrt{s}$, normalized to the energy density at $\sqrt{s} =2.76\TeV$.  The $\pt$ threshold used for jets is 10\GeVc at all centre-of-mass energies. Corrected results are compared to MC models used in cosmic ray physics and to \textsc{cascade} and \textsc{dipsy}.  Statistical uncertainties are smaller than the marker size, while the grey band represents the statistical and systematic uncertainties added in quadrature.}
\label{fig:ratiovss_cr}
\end{figure}

Figures~\ref{fig:ratiovss} and \ref{fig:ratiovss_cr} present the increase of the energy density deposited in the range $-6.6 < \eta < -5.2$ as a function of $\sqrt{s}$, normalized to the energy density at $\sqrt{s} = 2.76\TeV$, for both inclusive events and for events with a central charged-particle jet.  The $\sqrt{s} = 2.76$\TeV data are taken as a normalization point because this minimizes the statistical and systematic uncertainties. The $\pt$ threshold for jets is 10\GeVc at all centre-of-mass energies.  Since this is well within the plateau region, the energy density does not change significantly as a function of the actual value of the threshold.  Both figures again show the same data points, but compared to different models.


None of the \PYTHIA or \HERWIG{}++ models describe the increase with $\sqrt{s}$ seen in data.  For inclusive events the predictions differ little and they all underestimate the increase from $\sqrt{s} = 2.76$ to 7\TeV (by up to ${\sim}20\%$ for \HERWIG{}++ 2.5).  In this event class, the contribution of the underlying event is expected to be small.  For events with central charged-particle jets, the predictions vary more widely.  Indeed, for this event class the description of the underlying event in various tunes is expected to differ. None of the tunes give a satisfactory description, with \PYTHIA6 D6T and \PYTHIA8 4C being closest to the data and \HERWIG{}++ 2.5 underestimating the increase from 2.76 to 7\TeV by $\sim 25\%$.  The  \textsc{cascade} and \textsc{dipsy} generators also show a slower increase of the forward energy density with $\sqrt{s}$ than observed in data.  Of the cosmic ray models, \textsc{QGSJet}II-03 gives a good description of data. The \textsc{EPOS} and \textsc{sybill} generators yield an increase with centre-of-mass energy that is lower than that in the data by 10--15\%.

The results presented in this paper show that the MPI model, as implemented in \PYTHIA, and tuned to central inclusive and underlying event data, is capable of describing the $\pt$ dependence of the forward energy density. This is an important consistency check of the MPI model.   Models inspired by BFKL or CCFM parton dynamics do not describe the $\pt$ dependence of the data. Hence, contributions which go beyond what is presently implemented in the models seem to be mandatory.  Models used for cosmic rays studies, which include MPI and saturation effects via multi-pomeron interactions work well. {The \PYTHIA6 model with tune D6T describes the $\sqrt{s}$ dependence well, but only by invoking too large an amount of MPIs, as can be concluded from the $\pt$ dependence.

\section{Summary}
\label{sec:conclusion}

A study of the underlying event at forward pseudorapidity ($-6.6 < \eta < -5.2$) has been performed with a novel observable.  The energy density per unit of pseudorapidity has been measured at $\sqrt{s} =0.9$, 2.76, and 7\TeV, for events with a central charged-particle jet, relative to the energy density for inclusive events.  This hard-to-inclusive forward energy ratio has been studied as a function of the jet transverse momentum $\pt$.  In addition, the relative increase of the energy density as a function of the centre-of-mass energy has been measured for both inclusive events and events with a central charged-particle jet.  All results have been corrected to stable-particle level.

These results complement those obtained from studies of the underlying event at central rapidity \cite{cms2010,atlas2011,atlas2011-2,cms2011} because the large $\eta$ separation from the central hard scattering system yields a different sensitivity to the relative contributions of parton showers and multiple-parton interactions.  These data can thus be used to tune the underlying event parameters in a way which is complementary to that possible with central-rapidity data.

The data exhibit the typical underlying event behaviour characterized by a rapid change of the energy density at small charged-particle jet $\pt$, followed by a plateau at larger $\pt$.
At $\sqrt{s} = 7$\TeV, the relative energy density increases with jet $\pt$, while at $\sqrt{s} = 0.9$\TeV, the energy density decreases with increasing jet $\pt$. At this center-of-mass energy,
the hard-to-inclusive forward energy ratio drops below 1, which suggests that the energy of the proton remnant is depleted in events with a central charged-particle jet. Data at $\sqrt{s} = 2.76$\TeV
exhibit an intermediate behaviour and are characterized by an approximately constant energy density as a function of the jet $\pt$.

Models based on multiple-parton interactions suggest that the latter only make a limited contribution to the forward energy density in inclusive events. In contrast, collisions with a small impact
parameter, characterized by the presence of a charged-particle jet, appear to give rise to a significant number of multiple-parton interactions. Above $\pt = 8$\GeVc,
the hard-to-inclusive forward energy ratio is roughly independent of $\pt$, indicating that the collisions are already central for this value of the jet $\pt$. Some Monte Carlo models are able
to describe the hard-to-inclusive forward energy ratio as a function of $\pt$; however, all models fail to reproduce the dependence on the centre-of-mass energy simultaneously for inclusive events
and for events with a central charged-particle jet.

\section*{Acknowledgements}

We would like to thank Leif L{\" o}nnblad for making the \textsc{dipsy} predictions available to the CMS collaboration.

We congratulate our colleagues in the CERN accelerator departments for the excellent performance of the LHC and thank the technical and administrative staffs at CERN and at other CMS institutes for their contributions to the success of the CMS effort. In addition, we gratefully acknowledge the computing centres and personnel of the Worldwide LHC Computing Grid for delivering so effectively the computing infrastructure essential to our analyses. Finally, we acknowledge the enduring support for the construction and operation of the LHC and the CMS detector provided by the following funding agencies: BMWF and FWF (Austria); FNRS and FWO (Belgium); CNPq, CAPES, FAPERJ, and FAPESP (Brazil); MEYS (Bulgaria); CERN; CAS, MoST, and NSFC (China); COLCIENCIAS (Colombia); MSES (Croatia); RPF (Cyprus); MoER, SF0690030s09 and ERDF (Estonia); Academy of Finland, MEC, and HIP (Finland); CEA and CNRS/IN2P3 (France); BMBF, DFG, and HGF (Germany); GSRT (Greece); OTKA and NKTH (Hungary); DAE and DST (India); IPM (Iran); SFI (Ireland); INFN (Italy); NRF and WCU (Republic of Korea); LAS (Lithuania); CINVESTAV, CONACYT, SEP, and UASLP-FAI (Mexico); MSI (New Zealand); PAEC (Pakistan); MSHE and NSC (Poland); FCT (Portugal); JINR (Armenia, Belarus, Georgia, Ukraine, Uzbekistan); MON, RosAtom, RAS and RFBR (Russia); MSTD (Serbia); SEIDI and CPAN (Spain); Swiss Funding Agencies (Switzerland); NSC (Taipei); ThEPCenter, IPST and NSTDA (Thailand); TUBITAK and TAEK (Turkey); NASU (Ukraine); STFC (United Kingdom); DOE and NSF (USA). Individuals have received support from the A.\@ G.\@ Leventis Foundation, the Helmholtz Association, the Russian Foundation for Basic Research, the Russian Federation Presidential Grants N1456.2008.2, N4142.2010.2 and N3920.2012.2, the Russian Ministry of Education and Science, and the Belgian Federal Science Policy Office.

\bibliography{auto_generated}   

\cleardoublepage \appendix\section{The CMS Collaboration \label{app:collab}}\begin{sloppypar}\hyphenpenalty=5000\widowpenalty=500\clubpenalty=5000\input{FWD-11-003-authorlist.tex}\end{sloppypar}
\end{document}

%% file: FWD-11-003-authorlist.tex
\textbf{Yerevan Physics Institute,  Yerevan,  Armenia}\\*[0pt]
S.~Chatrchyan, V.~Khachatryan, A.M.~Sirunyan, A.~Tumasyan
\vskip\cmsinstskip
\textbf{Institut f\"{u}r Hochenergiephysik der OeAW,  Wien,  Austria}\\*[0pt]
W.~Adam, E.~Aguilo, T.~Bergauer, M.~Dragicevic, J.~Er\"{o}, C.~Fabjan\cmsAuthorMark{1}, M.~Friedl, R.~Fr\"{u}hwirth\cmsAuthorMark{1}, V.M.~Ghete, N.~H\"{o}rmann, J.~Hrubec, M.~Jeitler\cmsAuthorMark{1}, W.~Kiesenhofer, V.~Kn\"{u}nz, M.~Krammer\cmsAuthorMark{1}, I.~Kr\"{a}tschmer, D.~Liko, I.~Mikulec, M.~Pernicka$^{\textrm{\dag}}$, D.~Rabady\cmsAuthorMark{2}, B.~Rahbaran, C.~Rohringer, H.~Rohringer, R.~Sch\"{o}fbeck, J.~Strauss, A.~Taurok, W.~Waltenberger, C.-E.~Wulz\cmsAuthorMark{1}
% ...
\vskip\cmsinstskip
\dag:~Deceased\\
1:~~Also at Vienna University of Technology, Vienna, Austria\\
2:~~Also at CERN, European Organization for Nuclear Research, Geneva, Switzerland\\
% ...
